\documentclass{article} 
\usepackage{iclr2025_conference,times}
\usepackage{graphicx}

\usepackage{amsmath,amsfonts,bm}

\def\eqref#1{equation~\ref{#1}}

\def\1{\bm{1}}

\DeclareMathAlphabet{\mathsfit}{\encodingdefault}{\sfdefault}{m}{sl}
\SetMathAlphabet{\mathsfit}{bold}{\encodingdefault}{\sfdefault}{bx}{n}

\usepackage{hyperref}
\usepackage{url}

\title{Multi-modal single-cell foundation models via dynamic token adaptation}

\author{Wenmin Zhao, Ana Solaguren-Beascoa, Grant Neilson, Louwai Muhammed, Liisi Laaniste, \\ \textbf{and Sera Aylin Cakiroglu} \\ 
Cosyne Therapeutics \\ 
London, UK \\
\texttt{\{millie, ana, grant, louwai, liisi, aylin\}@cosyne.com} 
}
\iclrfinalcopy 

\begin{document}

\maketitle

\begin{abstract}
Recent advances in applying deep learning in genomics include DNA-language and single-cell foundation models. However, these models take only one data type as input. We introduce dynamic token adaptation and demonstrate how it allows combining these models to predict gene regulation at single-cell level in different genetic contexts.  Although the method is generalisable, we focus on an illustrative example by training an adapter from DNA-sequence embeddings to a single-cell foundation model's token embedding space. As qualitative evaluation, we assess the impact of DNA sequence changes on the model’s learned gene regulatory networks by mutating the transcriptional start site of the transcription factor \textit{GATA4} \textit{in silico}, observing predicted expression changes in its target genes in fetal cardiomyocytes.
\end{abstract}

\section{Introduction}
There have been rapid advances in training single-cell foundation models on large single-cell RNA sequencing (scRNA-seq) datasets \citep{theodoris_transfer_2023, yang_scbert_2022, cui_scgpt_2023}. These models represent each gene in a single-cell transcriptome as an input text token and do not integrate genetic information, making it hard to interpret gene expression predictions under genetic changes. On the other hand, DNA language models have been trained to predict epigenetic signals from the DNA sequence \citep{kelley_sequential_2018, kelley_cross-species_2020, avsec_base-resolution_2021, avsec_effective_2021}, and can be fine-tuned to predict the expression values of individual genes across cells in a scRNA-seq dataset \citep{schwessinger_single-cell_2023}. 
However, these models do not take cell-level co-regulation into account when predicting a gene's expression, instead, they focus on predicting epigenetic signals or the expression for each gene separately.

Our method of combining the modelling of both DNA sequences and single-cell transcriptomics data is inspired by unified embedding architectures for multi-modal large language models (LLMs), which convert an image into embedding vectors as a set of special tokens that are prepended to the input text tokens \citep{cho_unifying_2021}. Similar approaches using an adapter to provide additional information to the transcriptome have recently been applied to single-cell models \citep{maleki_efficient_2025, levine_cell2sentence_2023}, however, these methods have been restricted to a few additional tokens encoding a single entity (e.g. cell-type, disease state, molecule of drug treatment). 

In this paper, we propose extending the approach to all tokens in the input to allow their embeddings to flexibly encode additional information from a different modality that may change between data samples, which we call dynamic token adaptation (DTA). 
As an application, we introduce Bio-DTA, a novel multi-modal model that learns from single-cell transcriptomes and DNA sequences jointly. Finally, we demonstrate that the model has learned dynamic co-regulation by assessing the impact of genetic changes to the DNA sequence of the transcription factor \textit{GATA4} \textit{in silico} on the model's predictions for its targets.

\section{Methods}

\paragraph{Training data}
We downloaded scRNA-seq data from $33,364,242$ unique cells across $265$ datasets in the census dataset (version 2023-07-25) from the \href{https://cellxgene.cziscience.com/}{ CellXGene data portal } \citep{czi_single-cell_biology_program_cz_2023}. 
Data processing followed \citet{theodoris_transfer_2023}, representing each single-cell transcriptome as a sequence of gene names of maximum length $2,048$ ordered by their median-normalised expression. We excluded cancer cells and cells with $<500$ expressed genes. Transcriptional start sites (TSSs) of all protein coding genes were obtained from Ensembl (GRCh38.108). For each gene, $196,608$bp around the TSS of the reference genome was inputted to Enformer to compute the mean embedding of dimension $3,072$ across the positions of the pooling convolutional filters. The Enformer checkpoint was obtained from \href{https://github.com/lucidrains/enformer-pytorch}{https://github.com/lucidrains/enformer-pytorch}.

\begin{figure}[!tbp]
        \centering
        \includegraphics[width=0.4\textwidth]{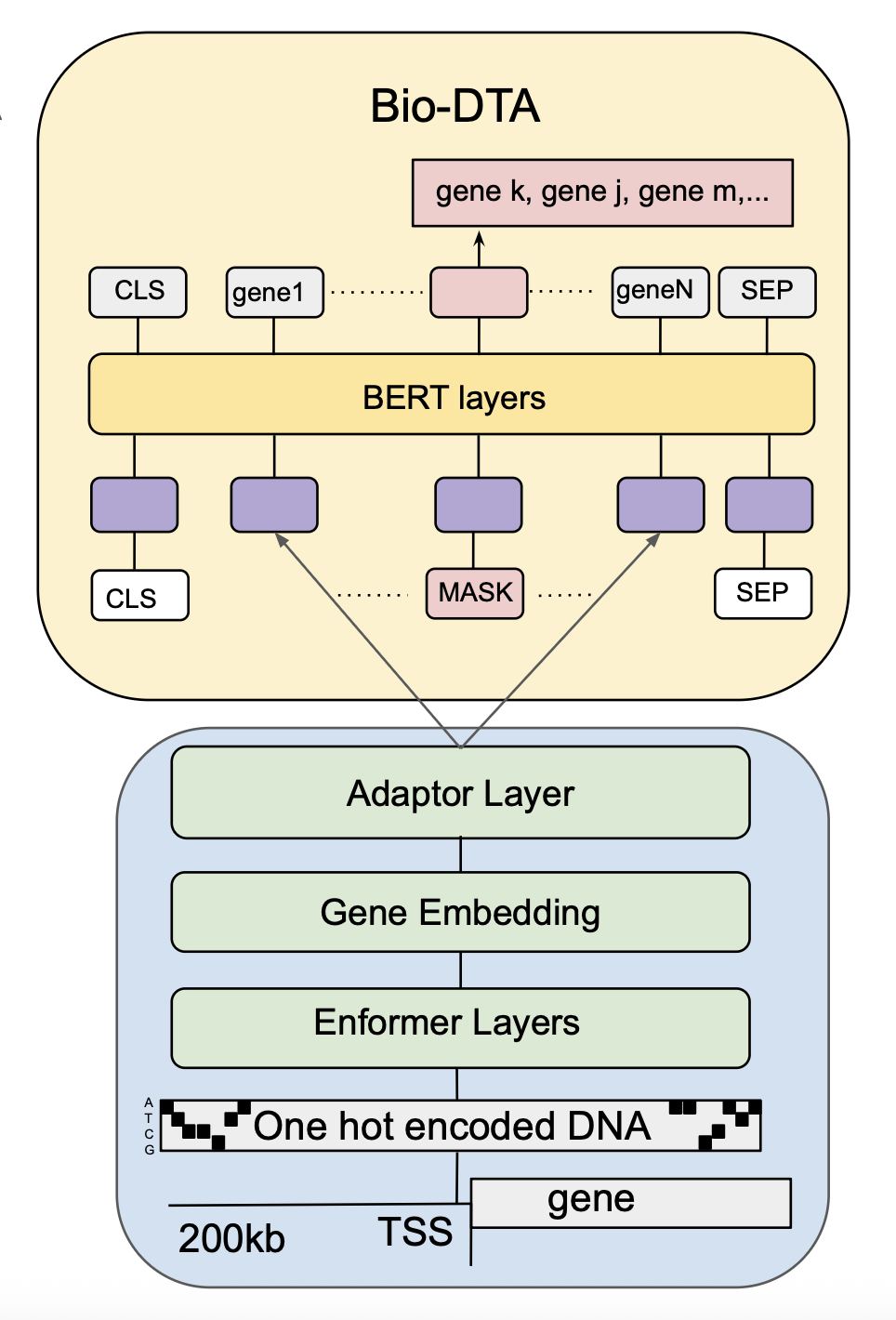}
       \caption{
        Schematic of Bio-DTA. Fixed token embeddings are replaced with a projection of aggregated Enformer embeddings of the gene’s DNA sequence.}
    \label{fig:main_schema}
\end{figure}

\paragraph{Model architecture and training}
Bio-DTA combines a DNA language model with a single-cell foundation model via token adapters. Although our method is flexible and can accommodate other architectures, in the experiments presented here, the single-cell foundation model is based on a bidirectional transformer encoder-only architecture (BERT). The model receives a single-cell transcriptome of length $2,048$ as an ordered sequence of gene names as input, which are mapped to integer identifiers called token IDs \citep{theodoris_transfer_2023}.  In a usual BERT model, each token ID is mapped to a unique and trainable embedding vector that forms the input to the transformer encoder \citep{devlin_bert_2019}. In contrast, Bio-DTA (Figure \ref{fig:main_schema}) projects Enformer's aggregated embeddings for each input gene to the token embedding size using an adapter layer (e.g. a multilayer perceptron followed by a softplus activation). This forms the input to the transformer encoder capturing genetic information. Here, a gene's token embedding is not unique and may change if the gene's input DNA sequence changes, and the same adapter is used for all the genes. Bio-DTA outputs the token IDs of the input gene names according to the single-cell foundation model. 

Bio-DTA was trained end-to-end with a masked language modelling task (masking $15\%$ of input tokens) for three epochs. 
For more implementation details as well as hyperparameters used for the final model and training procedure see Appendix \ref{implementation_details}.

\section{In silico mutagenesis reveals learned connection \textbf{dynamics} in Bio-DTA}

\begin{figure}[h]
    \centering
    \includegraphics[width=\linewidth]{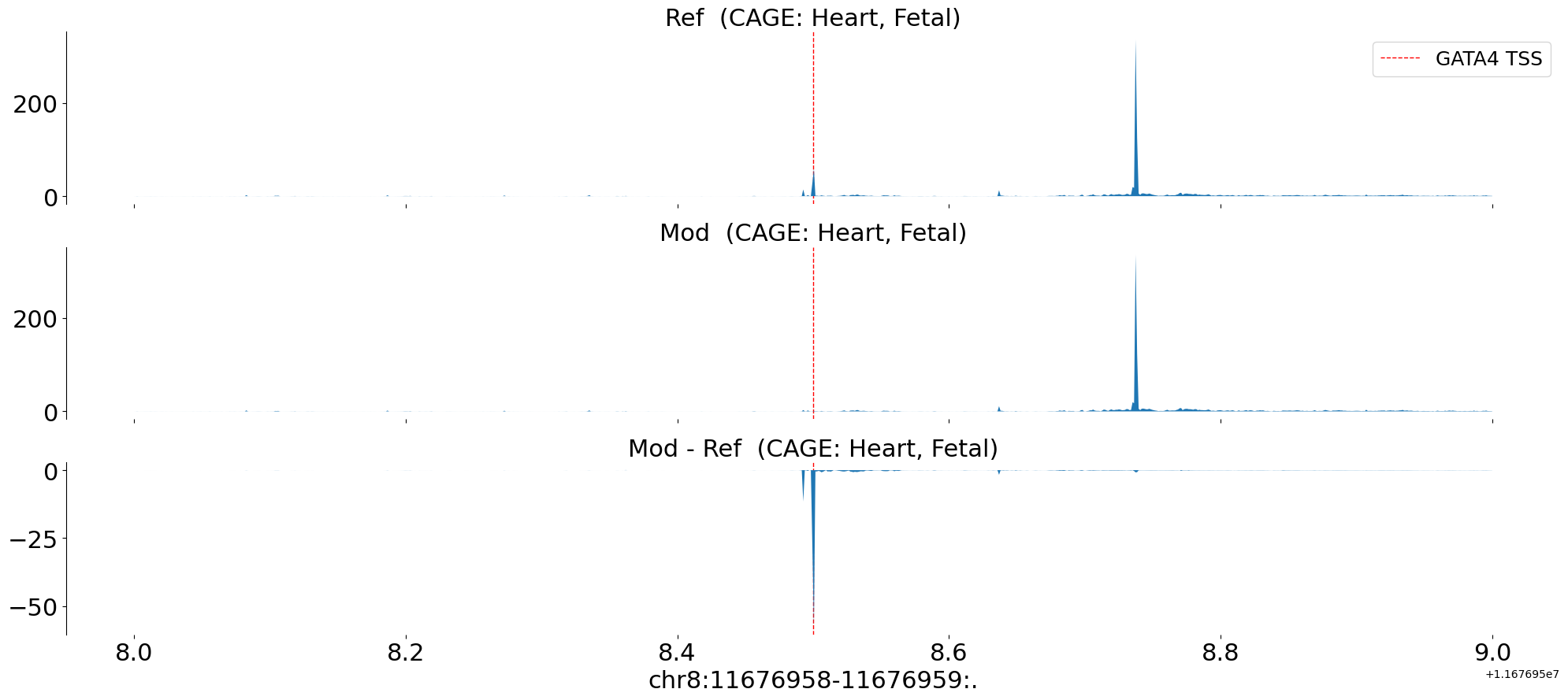}
    \caption{Enformer predicted Cap analysis of gene expression (CAGE) tracks of gene expression centered around the \textit{GATA4} TSS for the reference genome (top), \textit{in silico} mutated sequence (random sequence replacing $100$bp around TSS) (middle), and the difference in predictions between the \textit{in silico} mutated and reference sequence showing a large decrease around the TSS in the middle of the track (bottom). }
    \label{fig:figure_insilico_perturbation_GATA4_expression}
\end{figure}

To assess how changes in the input DNA sequence of one gene affects learned co-regulation networks, we followed \citet{theodoris_transfer_2023} and focussed on \textit{GATA4} and \textit{TBX5}, two known congenital heart disease genes, which are co-expressed during cardiac morphogenesis, physically interact and have co-bound targets \citep{misra_disruption_2014}. We obtained $103$ transcriptomes from fetal cardiomyocytes expressing \textit{GATA4} \citep{knight-schrijver_single-cell_2022}. We then introduced random mutations \textit{in silico} in $100$bp around the \textit{GATA4} TSS, which reduced Enformer's predicted expression (Figure \ref{fig:figure_insilico_perturbation_GATA4_expression}). For each single-cell transcriptome in the dataset, we performed a forward pass and extracted the contextualised embedding for each input gene from
the penultimate layer of Bio-DTA, once when using the unchanged \textit{GATA4} embedding, and once when using the mutated embedding as input for \textit{GATA4}.  As the retained gene embeddings are from deeper layers of Bio-DTA's BERT encoder, they may capture information about other co-regulated genes in the input sequence. To test if these depend on genetic changes, we calculated the cosine similarity between the gene embeddings before and after \textit{GATA4}'s {\textit{in silico}} mutagenesis for each transcriptome.  Indeed, the embeddings of experimentally identified GATA4 and TBX5 targets (defined based on ChIP-seq data in \citet{theodoris_transfer_2023}) had a significant drop in cosine similarity compared with the remainder of the genome ($p< 0.05$, Wilcoxon test), while embeddings of housekeeping genes remained stable (Table \ref{tab:gene_significance}).

\begin{table}[h]
    \centering
    \renewcommand{\arraystretch}{1.5}
    \begin{tabular}{l|c|c}
        & P-value & FDR adjusted P-value \\
        \hline
        GATA4 direct & 6.42e-10 & \bf{3.85e-09} \\
        GATA4 indirect & 3.41e-07 & \bf{5.11e-07} \\
        GATA4 \& TBX5 combination & 1.51e-09 & \bf{4.52e-09}  \\
        TBX5 direct & 2.64e-09 & \bf{5.27e-09} \\
        TBX5 indirect & 5.75e-07 & \bf{6.90e-07} \\
        housekeeping genes & 7.24e-01 & 7.24e-01  \\
    \end{tabular}
    \caption{P-values of a Wilcoxon test comparing the cosine similarities of the gene embeddings before and after \textit{in silico} mutagenesis of \textit{GATA4} in a target group with those of the remainder of the genome not in any of the target groups. Significant FDR-adjusted p-values at a significance threshold of $0.05$ are highlighted in bold. Target groups as indicated were obtained from \citet{theodoris_transfer_2023}.}
    \label{tab:gene_significance}
\end{table}

Next, we selected the $50$ genes with the largest changes in their embeddings and assessed the model's capability to identify experimentally derived targets compared with a random gene set of the same size. To compare the performance of Bio-DTA with Geneformer, we perform \textit{in silico} deletion of GATA4 by removing it from the input sequence as described in \cite{theodoris_transfer_2023}. We also train a BERT model similar to Geneformer with the same parameters as Bio-DTA that does not use adapters and instead uses the gene names as input sequence on the same training data. We perform \textit{in silico} deletion of GATA4 for this model as described for Geneformer.  

\begin{figure}[!tbp]
        \centering
        \includegraphics[width=0.45\textwidth]{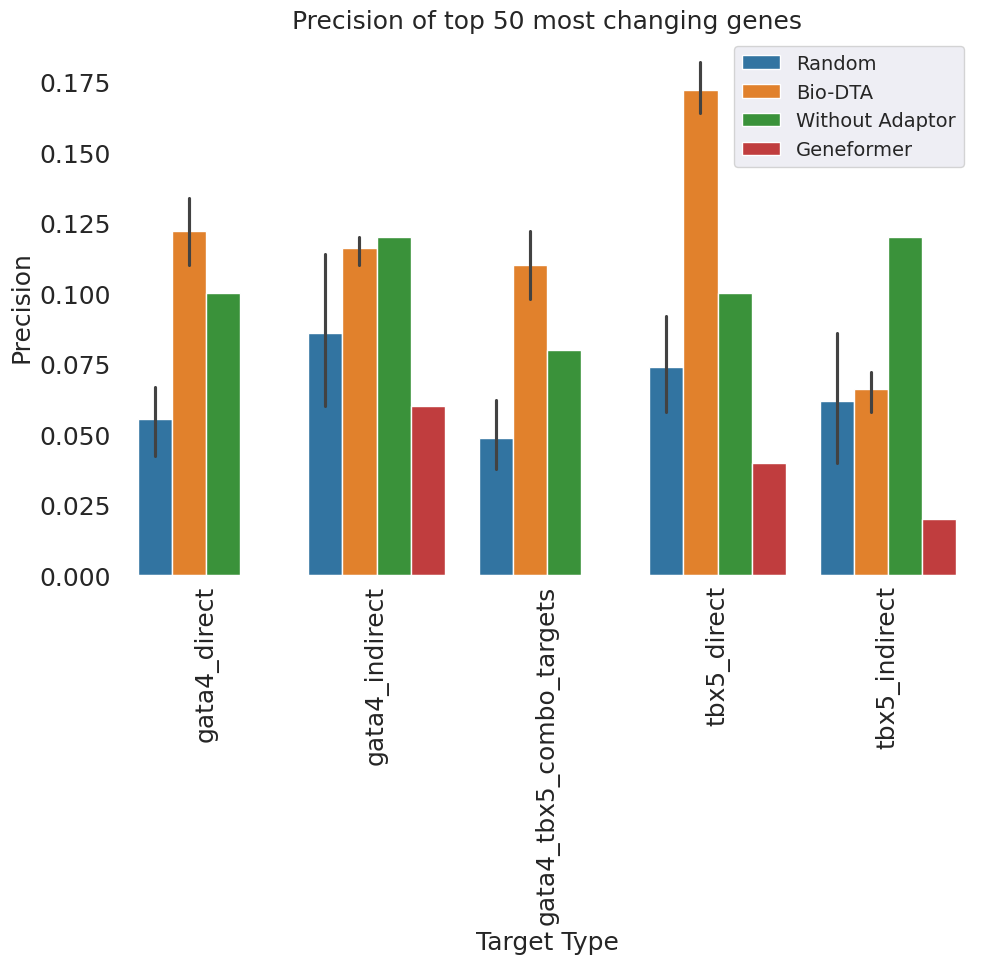}
        \includegraphics[width=0.45\textwidth]{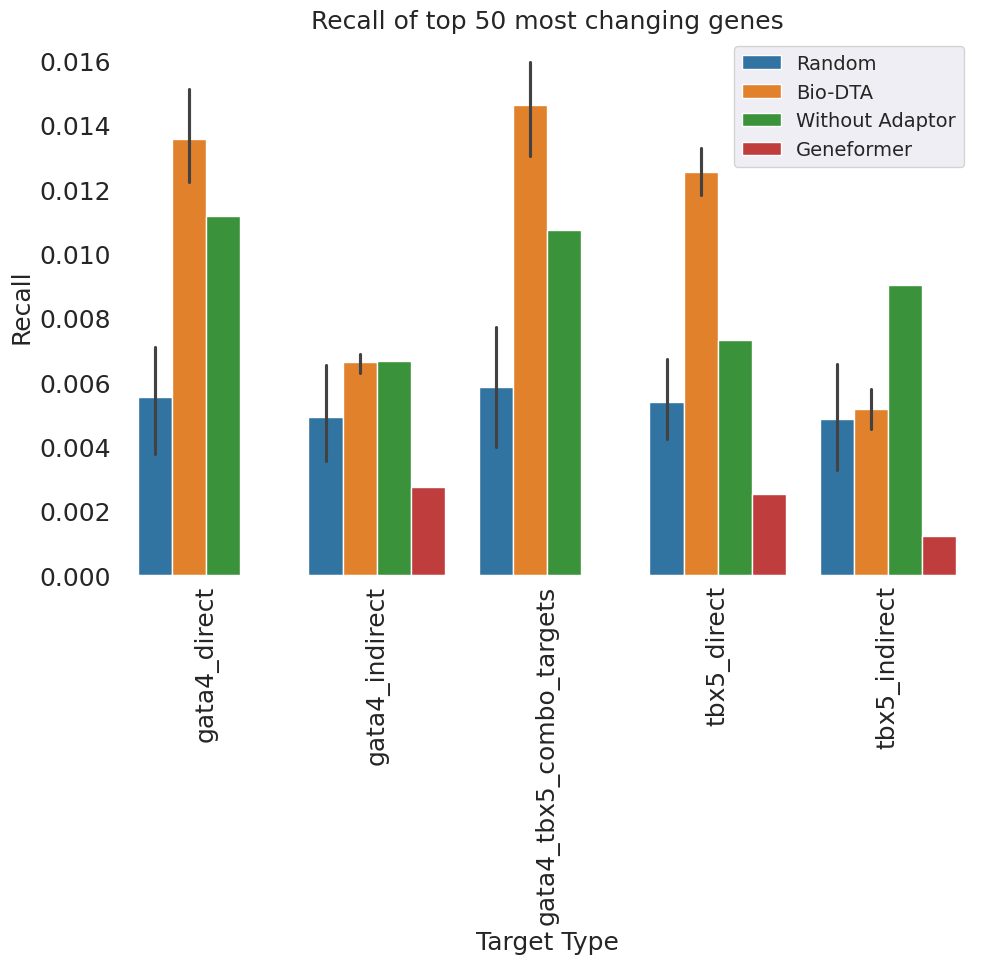}
       \caption{
        Barplots of precision and recall of putative GATA4 and TBX5 targets (orange) based on contextualised embedding changes under \textit{in silico} mutagenesis of \textit{GATA4} in fetal cardiomyocytes, comparing with the in-silico deletion of \textit{GATA4} in the BERT model without an adapter (green), Geneformer (red) and size-matched random samples (blue). Target group annotations based on ChIP-seq data were obtained from \citep{theodoris_transfer_2023}.}
    \label{fig:barplots}
\end{figure}

Figure \ref{fig:barplots} shows the precision and recall of true targets for each of the models in each group. Bio-DTA performs best on both metrics for direct GATA4 and TBX5 and their co-bound targets, and is comparable to the adapter-free model on indirect GATA4 targets. However, it is outperformed by the adapter-free model on indirect TBX5 targets.   Across all groups and metrics, Geneformer is outperformed by both models and sometimes the random gene set baseline. In particular, it can not recover any experimentally verified GATA4 targets. A similar trend can be seen using the top $100$ genes (Figure \ref{fig:barplots_top100}).
Taken together, our results demonstrate how the dynamic token embeddings allow the model to be sensitive to small genetic changes and their impact on its learned co-regulation networks.

\section{Conclusion}
\label{others}
We introduced dynamic token adaption to project embeddings of a different data modality into the text token embeddings of a foundation model. As an application, we presented Bio-DTA, a multi-modal model that combines a DNA language model with a single-cell transcriptome model. 

We showed that Bio-DTA learns the impact of small genetic changes on co-regulation networks. We evaluated the model's ability to capture gene co-regulation by performing \textit{in silico} mutagenesis of the GATA4 transcription factor and observing changes in the contextualized embeddings of target genes in fetal cardiomyocytes. 
Next, we showed that Bio-DTA successfully returns experimentally verified targets amongst the $50$ genes with the largest changes in their embeddings. 
We compared these results with an adapter-free model that was trained on the same data with the same hyperparameters. 
We showed that Bio-DTA outperforms the adapter-free model on direct GATA4 and TBX5 targets. This may be due to the method of \textit{in silico} deletion completely removing GATA4 from the input sequence, which changes the entire input transcriptome and removes learned connections between GATA4 input embeddings and its targets. In contrast, Bio-DTA uses small changes to the embedding of GATA4 to signal reduced expression without changing the overall input and allows the assessment of these changes on learned connections. On the other hand, the adapter-free model outperforms Bio-DTA on indirect targets. This may be due to the complete removal of GATA4's input during the \textit{in silico} deletion, resulting in a much larger change to the input, which may influence the contextualised embeddings more in the adapter-free model. 

In future work, we plan to extend our evaluation beyond the GATA4 case study by exploring additional transcription factors and cell types to validate the generalisability of Bio-DTA further. However, identifying suitable benchmark cases for such evaluation remains a significant challenge. While there are many examples of the impact of genetic changes on one gene's expression (for example, eQTLs), high-quality, well-characterized instances where both the genetic perturbation and its downstream co-regulation impact are experimentally validated — especially in a cell-type-specific context — are rare. This limits the availability of ground truth data against which to benchmark model predictions. Furthermore, we will expand our evaluations to more complex genetic variations such as SNPs, indels, and structural variants which introduce additional biological layers that will provide a more comprehensive evaluation of the model.

In the work presented here, the reference genome was used as input to Enformer and \textit{in silico} mutagenesis was performed as a zero-shot approach. DNA language models trained on the reference genome such as Enformer struggle to reliably predict the direction of eQTLs and the expression variation for different individuals \citep{huang_personal_2023-1, sasse_benchmarking_2023}. For a personalised medicine approach, future work will also include fine-tuning Enformer on different genomes with allelic information as described in \citet{drusinsky_deep-learning_2024} to provide more nuanced DNA sequence embeddings.


Other applications of DTA will include projecting embeddings of the gene's RNA isoforms or amino acid sequence into the token embedding space. While we mapped the DNA sequence to one token in the presented experiments, we will also evaluate using several tokens per input gene representing the DNA sequence to allow the encoding of larger genetic contexts.

\bibliography{iclr2025_conference}
\bibliographystyle{iclr2025_conference}

\appendix
\counterwithin{figure}{section}

\section{Appendix}
\subsection{Implementation}\label{implementation_details}
We used the same BERT-based architecture as in \citet{theodoris_transfer_2023} with $6$ transformer layers with input size $2,048$, embedding dimensions $256$, $4$ attention heads per layer and feed-forward size of $512$. The adapter is a single MLP layer with a Softplus activation. Hyperparameters were chosen to allow for distributed learning: max learning rate, $1 \times 10^{–3}$ scaled by the number of GPUs; a learning scheduler, linear with warm-up ($10$k steps) and linear decay; Adam optimizer with weight decay parameter $0.001$. Training was distributed over $4$ GPUs in one node with minibatch size $11$ and $2$ gradient accumulation steps.

To speed up pretraining we used dynamic padding combined with a length-grouped sampler to minimise computation on padding. This sampler takes a randomly sampled megabatch and then orders minibatches by their length in descending order. Mini-batches are then dynamically padded, minimising the computation wasted on padding as sequences of similar lengths are grouped. The authors of Geneformer extended an existing version of this sampler from Huggingface transformers for the distributed case \citep{theodoris_transfer_2023, wolf_huggingfaces_2020}. However, neither of these samplers shuffle the mini-batches within the megabatch before passing them to the model, which resulted in a 60x-performance-drop of the trained model in our tests (in terms of training and test perplexity on smaller sample datasets) compared to model runs not employing the grouped-length batching. We implemented a shuffling of the mini batches which slightly diminishes the speed up during training.

For efficient data parallelisation across the GPUS, we used Deepspeed \citep{rasley_deepspeed_2020}. Overall, pre-training was achieved  in just over 7 days distributed across one node with four Nvidia A10G 24GB GPUs.

\begin{figure}[!tbp]
        \centering
        \includegraphics[width=0.45\textwidth]{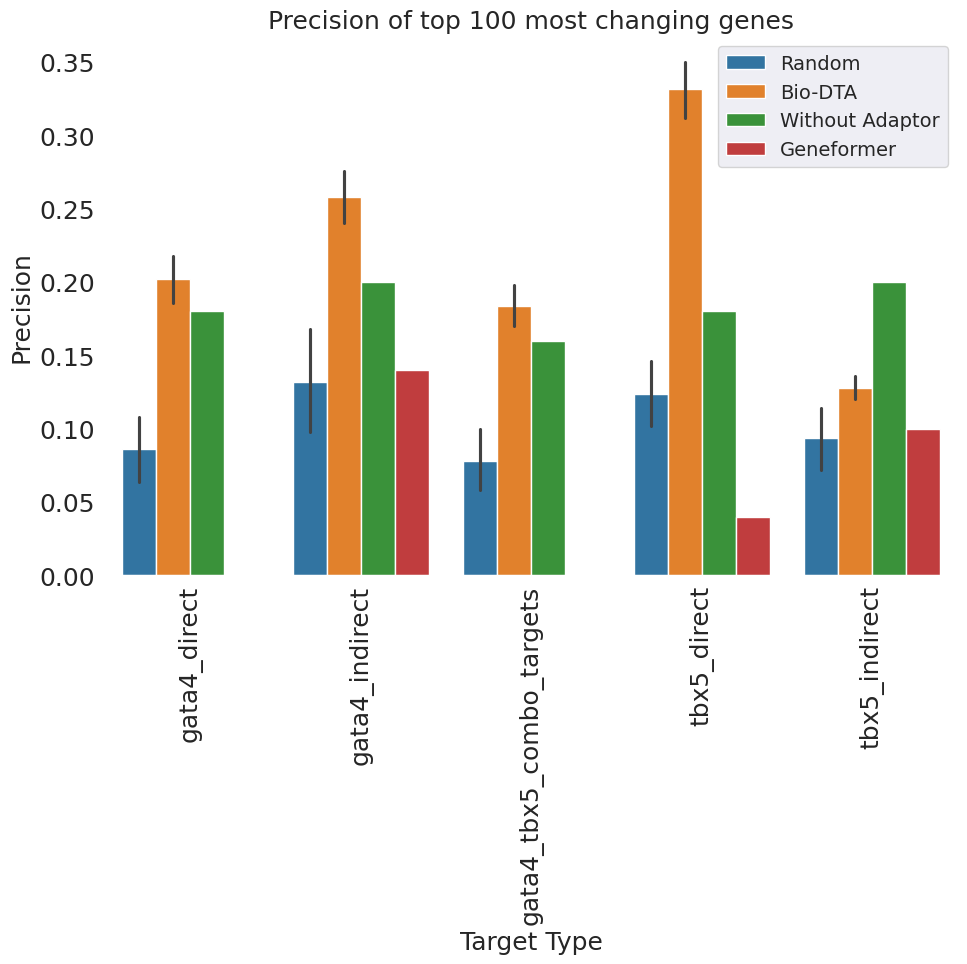}
        \includegraphics[width=0.45\textwidth]{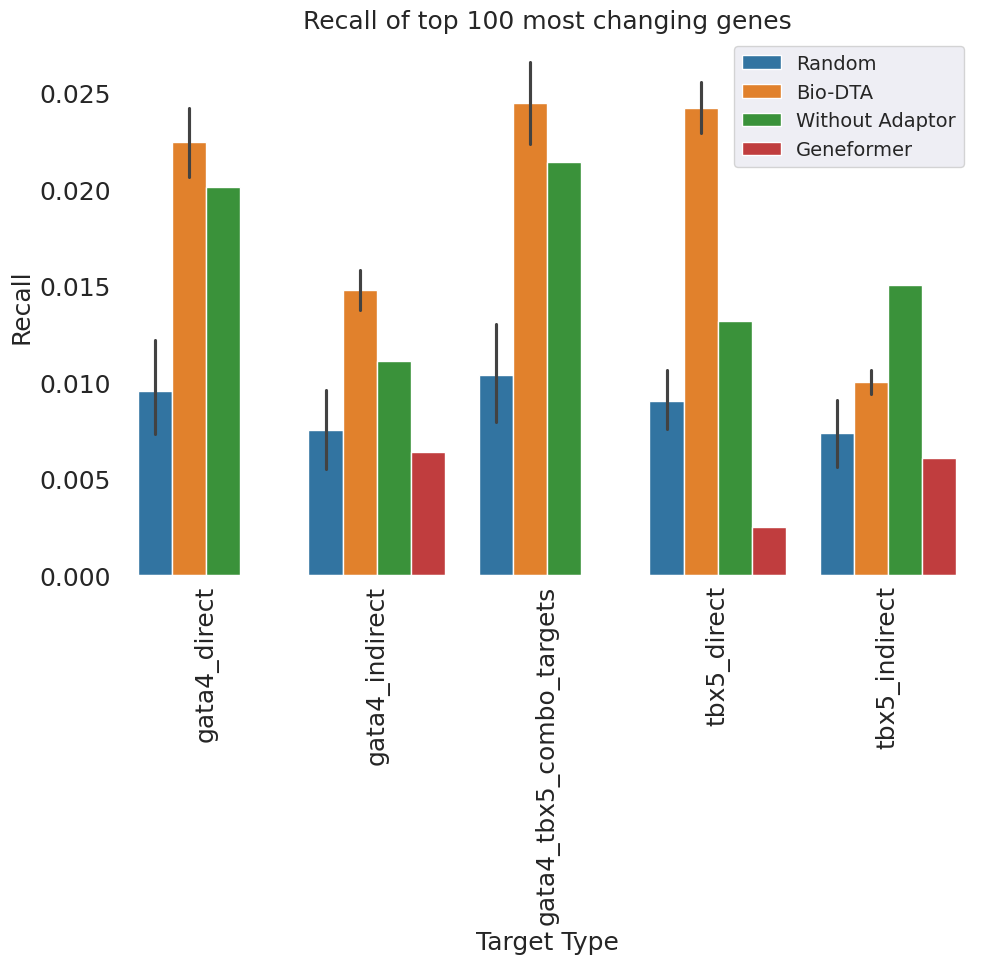}
       \caption{
        Barplots of precision and recall of putative GATA4 and TBX5 targets (orange) based on contextualised embedding changes under \textit{in silico} mutagenesis of \textit{GATA4} in fetal cardiomyocytes, comparing with the in-silico deletion of \textit{GATA4} in the BERT model without an adapter (green), Geneformer (red) and size-matched random samples (blue). Target group annotations based on ChIP-seq data were obtained from \citep{theodoris_transfer_2023}.}
    \label{fig:barplots_top100}
\end{figure}



\end{document}